\documentstyle[12pt,epsfig]{article}
\newcommand{\abs}[1]{\left|#1\right|}

\newcommand{\ket}[1]{\vert #1\rangle}
\newcommand{\braket}[2]{\langle #1\vert #2\rangle}
\newcommand{\eg}{e.\,g.\ }
\newcommand{\ie}{i.\,e.\ }

\renewcommand{\epsilon}{\varepsilon}

\begin{document}
\title{Does Indeterminism Give Rise to an Intrinsic Time Arrow?
\thanks{Submitted
to Studies in Hist. and Phil. of Mod. Phys.}}
\author{Shahar Dolev$^a$\thanks{email: shahar\_dolev@email.com},
Avshalom C. Elitzur$^a$\thanks{email: avshalom.elitzur@weizmann.ac.il},
Meir Hemmo$^b$\thanks{email: meir@research.haifa.ac.il}}
\maketitle

\begin{center}
$^a$Unit of Interdisciplinary Studies,
Bar-Ilan University, 52900,
Israel.

$^b$Philosophy Department, Haifa University, Haifa 31905
Israel.
\end{center}


\begin{abstract}
\noindent
In any physical theory that admits true indeterminism, the thermodynamic
arrow of time can arise regardless of the system's initial conditions.
Hence on such theories time's arrow emerges out of the basic physical interactions. The example of the GRW theory is studied in detail.

\end{abstract}

\section{Introduction}
\label{sec:intro}

Our experience suggests that the
universe's evolution is invariably {\em time-asymmetric},
that is, that events evolve one after another
from past to future.
If this intuition is correct, then time's apparent arrow should be
intrinsic to the fundamental dynamical laws describing our universe.
However, this intuitive impression is
contradicted by current physical theories, according to
which the fundamental dynamical laws of the universe are without
exception {\em time symmetric}, i.e., invariant
under time reversal.

In view of this time symmetry of physical laws,
most accounts in the foundations of physics
appeal to {\em statistical} assumptions concerning certain
special boundary conditions in order to account for the apparent
arrow of time. There are, however a few attempts (see \eg Penrose \cite{Pen1979}) 
to account for this arrow of time by making an appeal to a new
physics which would explicitly include within the dynamical equations of
motion a time-asymmetric component.

Another unresolved question in the foundations of physics is
whether {\em determinism} holds at the
microscopic level. Here opinions are more evenly
divided. In quantum mechanics, for instance, some
interpretations postulate dynamical
equations of motion that are completely deterministic
(\eg Bohm's pilot-wave \cite{Bohm52} and Everett's many
worlds interpretation \cite{Everett57}), whereas other interpretations,
following early ideas
of von Neumann \cite{Von1932} and Dirac \cite{Dir1939}, attempt to
rewrite the
fundamental quantum mechanical laws by proposing {\em
indeterministic} equations of motion
(\eg the collapse theory of Ghirardi, Rimini, and Weber
(GRW) \cite{GRW1986} and Penrose's \cite{Pen1994} hypothesis of
gravitational collapse). Genuine
stochastic dynamics has also been invoked in other fields of physics,
\eg black-hole thermodynamics (Hawking and Penrose \cite{HP1996}) and
general relativity (Earman \cite{Ear1995}).

These two questions,
that is, the questions of whether the
fundamental dynamical laws contain an (as yet unknown) intrinsic
time arrow, and whether or not they are genuinely
deterministic, went on nearly
oblivious to one another. In this paper we would like to study their
bearings on one another in detail.

The structure of this paper is as follows. In section \ref{sec:uncert}
we
discuss three types of uncertainty and their physical manifestations. In
section \ref{sec:fail} we show the fundamental affinity between
indeterminism and entropy. In section \ref{sec:grw} we briefly review
the GRW theory, then
in section \ref{sec:alb} we review Albert's
method of deriving the thermodynamic arrow from quantum collapse. Finally, we
discuss
the broader connection between the stochastic dynamics and the
arrow of time.

\section{Three Uncertainties and their Physical Manifestations}
\label{sec:uncert}

Let us begin with some theoretical considerations of uncertainty, reversibility, and determinism. We will use the term ``uncertainty" to denote the strong, ontological sense of the term. In
other words, we do not refer to the observer's mere ignorance but to the real
absence of any strict cause-effect relations between two event. We make
use of the notion of causation here in an intuitive way without
appealing to any specific theory of causation. 
An operational definition for such uncertainty is as follows:

{\em If event {\rm A} is uncertain and event {\rm B} is its cause
(effect), then, repeating the process in the forward (backward) time
direction will not always reproduce {\rm A} from {\rm B}.}

Such an uncertainty can take one out of three forms, which we shall
denote by ``$\bf V$," ``$\bf\Lambda$," and ``$\bf X$," according to these
letters' shapes (as in Fig.~1). Strict determinism will be represented by $\bf
I$, since it has a linear topology: each initial event $C$ constitutes a
cause for only one effect $E$. Indeterministic theories can have one of
the  following topologies :``$\bf V$ uncertainty" arises when a certain
cause $C$ can give rise to more than one possible effect: $E_1$, $E_2$,
... . ``$\bf\Lambda$ uncertainty" is the inverse case where a system can
start at any one out of some initial states, $C_1$, $C_2$, ..., but
always reaches one final state $E$. ``$\bf X$ uncertainty" is a
combination of the former two: an event can have several initial causes,
$C_1$, $C_2$, ..., as well as several effects, $E_1$, $E_2$, ... .

\begin{figure}
\label{fig:uncert}
\begin{center}
\includegraphics[scale=0.6]{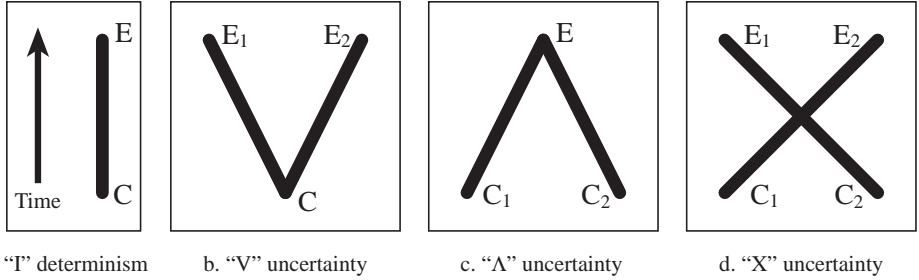}
\caption{Possible topologies for physical dynamics.}
\end{center}
\end{figure}

Note the following:
\begin{itemize}
\item Only $\bf \Lambda$ and $\bf V$ topologies are asymmetric under
time reversal, and as such have a ``built-in" microscopic time arrow.
This microscopic time arrow may, or may not, manifest itself at the
macro level, for instance in the form of a thermodynamic arrow
(in accordance with the second law of thermodynamics). However, any further discussion on the arrow of time becomes superfluous, since its origin is apparent.

\item $\bf V$ uncertainty is what we call ``indeterminism", since the
present state of the system cannot help predicting its future state. However,
such a topology is {\em reversible} since reversing the system from any
of the effects $E_1$, $E_2$, ..., will always yield the initial cause
$C$.

\item $\bf \Lambda$ uncertainty, on the other hand, is deterministic in
the forward time direction, but irreversible (in the sense just defined).

\item All but $\bf X$ uncertainty are not accessible to empirical
investigation. In $\bf I$ topology, for example, a system can be
prepared in such a way as to either increase or decrease its entropy,
depending only on the initial state. Hence, as Hawking \cite{Haw1976}
pointed out, we might be living in a universe that actually evolves from what
we call ``future" to our ``past", and we wouldn't notice, as our perceptual
mechanisms would be reversed accordingly. Similarly, since $\bf V$ and $\bf \Lambda$ uncertainties are mirror images of one
another, we could be living in a world having one topology which
nevertheless evolves from ``past" to ``future" (in some absolute sense),
or {\it vice versa}. Neither experience nor experiment can distinguish between these possibilities. 

\item Hence, only $\bf X$ is a testable hypothesis, which we shall study
in what follows.
\end{itemize}

\begin{figure}
\label{fig:exp}
\begin{center}
\includegraphics[scale=0.6]{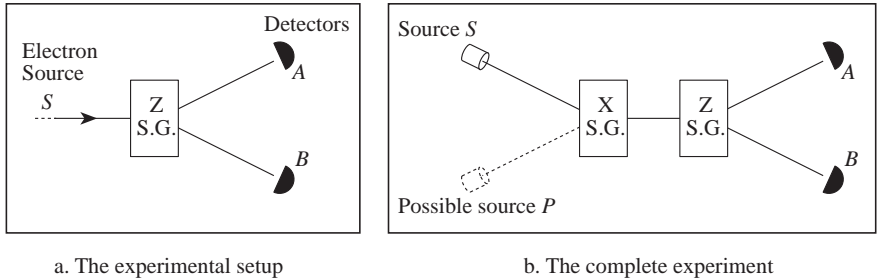}
\caption{How to view physical indeterminism.}
\end{center}
\end{figure}

Returning to physics, it is quantum mechanics where one would look for
examples for such uncertainties. We would like to state that uncertainty in physics always comes in the
$\bf X$ topology. For that purpose, consider the experimental setup in
Fig.~\ref{fig:exp}a. Here an electron source emits electrons in an $x$-spin up ($x+$)
state. These electrons pass through a Stern-Gerlach (S.G.)
magnet that splits their path according to their $z$-spin. Then the
electrons can hit, with a 50-50 probability, one of the two detectors:
$A$ for $z+$, or $B$ for $z-$. 

In a world of indeterministic dynamics (such as von Neumann's
``collapse", or GRW\footnote{What we say in what follows applies only
to the {\em discrete} model of the GRW theory. 
In the continuous model the
dynamics turns out also to be time irreversible, but for
different reasons related to the behavior of the tails of the
collapsed wavefunction (see footnote \ref{note:GRW} for more details).})
one would intuitively classify
its topology to be $\bf V$, such as in Fig.~1b: here one initial state $C$
(the electron at spin $x+$), gave rise to one out of two effects: $E_1$
(the electron hitting $A$ with spin $z+$), and $E_2$ (the electron hitting $B$
with spin $z-$). However, $\bf V$ topology must be strictly reversible, and this is {\em not}
the case here: reversing the operation of the detector that has clicked will cause the electron to reach the source in either a $z+$ or a $z-$
 state, in contrast to state $C$ where a
spin $x+$ electron was emitted.

In order to resolve this inconsistency, let us look at Fig.~\ref{fig:exp}b.
Here the apparatus is extended to the past to explain how one could make
an $x+$ spin source: one has actually used a generic electron source and passed
the output beam through an $X$-S.G. Only the $x+$ part was allowed to continue the
experiment. Now, reversing the experiment will, in 50\% of the cases,
return an $x+$ electron to the source $S$.

However, there is another possible past for the system. The
electron could have been emitted in a $X-$ spin state from source $P$,
and reach {\em exactly} the same probabilities for the final states
$E_1$, and $E_2$. This additional possible past now completes the
picture, since reversing from these final states might cause the
electron to be deflected down at the $x$-S.G. and reach source $P$.

Now we can see that in an indeterministic theory the dynamics can be
time reversed in the above sense provided there are 
{\em two} possible pasts - that is, an $\bf X$ topology. Note, however,
that the $\bf X$ might not be completely time-symmetric, that is $C_1$, and $C_2$ need
not be equal to $E_1$, and $E_2$.

Note further that in the case of collapse interpretations of quantum
mechanics such as the GRW theory (and von
Neumann's theory) the {\bf X} shape situation means that the statistical
results of measurements that may be 
carried out on a system at a given time (that is, of 
measurements of {\em commuting} observables) are
in principle insufficient in order to built up, by retrodiction, the initial
wavefunction of the system.\footnote{One can, however, built up the
wavefunction from the (counterfactual) statistical results of all 
possible measurements which would include non-commuting
measurements.} Of course, in quantum mechanics without collapse it is also
true that the actual statistics of results is insufficient to retrodict
with certainty the wavefunction, but in such theories it is also 
true that the overall state of the world after a measurement is described
by the uncollapsed wavefunction which obeys a completely
deterministic and time reversible dynamics (\ie the Schr\"odinger
equation).

Next we shall consider the bearing of these conclusions on the
origins
of the thermodynamic arrow of time.

\section{When Determinism Fails}
\label{sec:fail}

In most physical discussions, ``irreversibility" is used in the
technical, hence relative sense. Processes like milk being spilt or a
match being burnt are irreversible only in the practical sense because,
in principle, a sufficiently advanced technology can reverse them.
Indeed some processes considered irreversible by past standards are
becoming reversible today.

Notice, however, that when referring to ``sufficiently advanced
technology," one has in mind a kind of nanotechnology that can reverse,
precisely and simultaneously, the motions of a myriad of molecules, so
as to get spilt milk gathered anew in the jar and a match re-assembled
from charcoal, smoke, and thermal photons. Now, this sense of
reversibility rests on a highly non-trivial assumption that must no more
be left tacit. {\it To say that a process can be microscopically
reversed, one must profess absolute determinism.} It is only strict
determinism that guarantees that, once the momenta of all molecules of a
system are reversed, the system will return to its initial state.

\begin{figure}
\label{fig:sim1}
\centering
\includegraphics[scale=0.75]{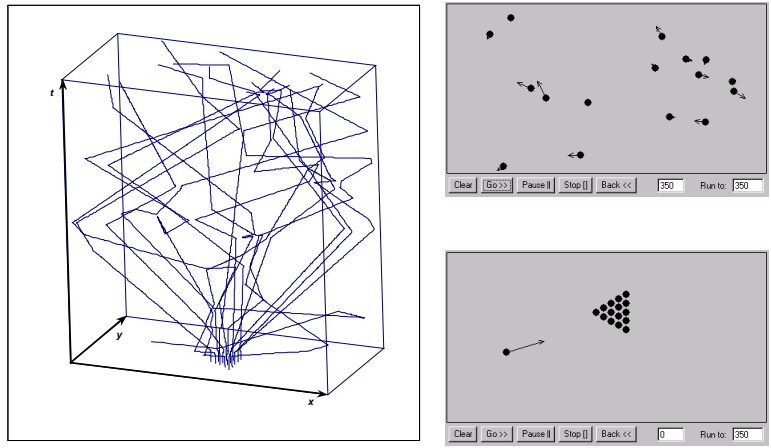}
\renewcommand{\thefigure}{3a}
\caption{A computer simulation of an entropy increasing process, with
the
initial and final states (right) and the entire process using a
spacetime
diagram (left). One billiard ball hits a group of ordered balls at rest,
dispersing them all over the table. After repeated collisions between
the
balls, the energy and momentum of the first ball is nearly equally
divided
between the balls.}
\end{figure}

\begin{figure}
\begin{center}
\includegraphics[scale=0.75]{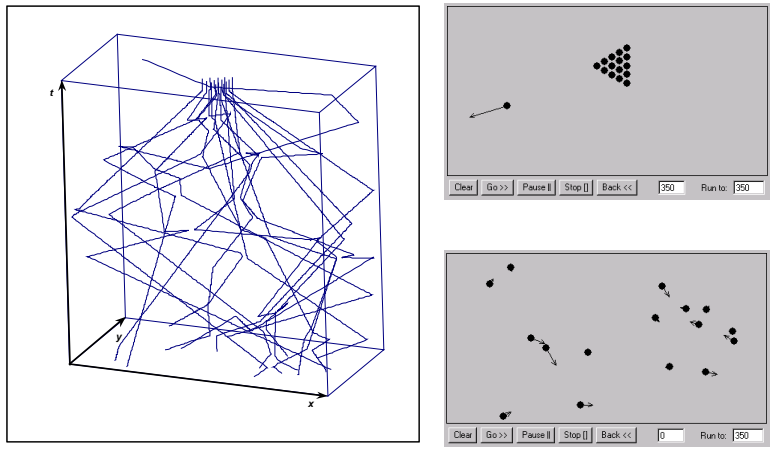}
\renewcommand{\thefigure}{3b}
\caption{The time-reversed process. All the momenta of the balls are
reversed
at t=350. Eventually, the initial ordered group is re-formed, as at t=0,
ejecting
back the first ball.}
\end{center}
\end{figure}

\begin{figure}
\begin{center}
\includegraphics[scale=0.8]{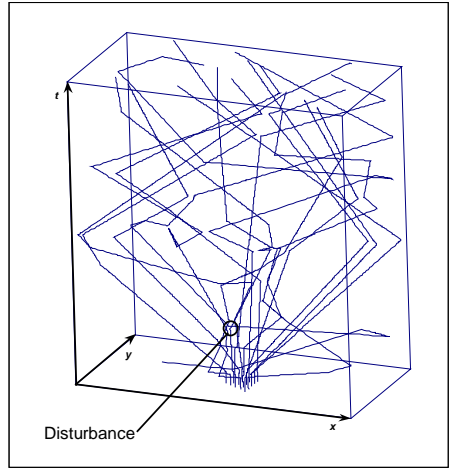}
\renewcommand{\thefigure}{4a}
\caption{The same simulation as in 3a, with a slight disturbance in the
trajectory of one ball (marked by the small circle). Entropy increase
seems to
be indistinguishable from that of 3a.}
\end{center}
\end{figure}

\begin{figure}
\begin{center}
\includegraphics[scale=0.8]{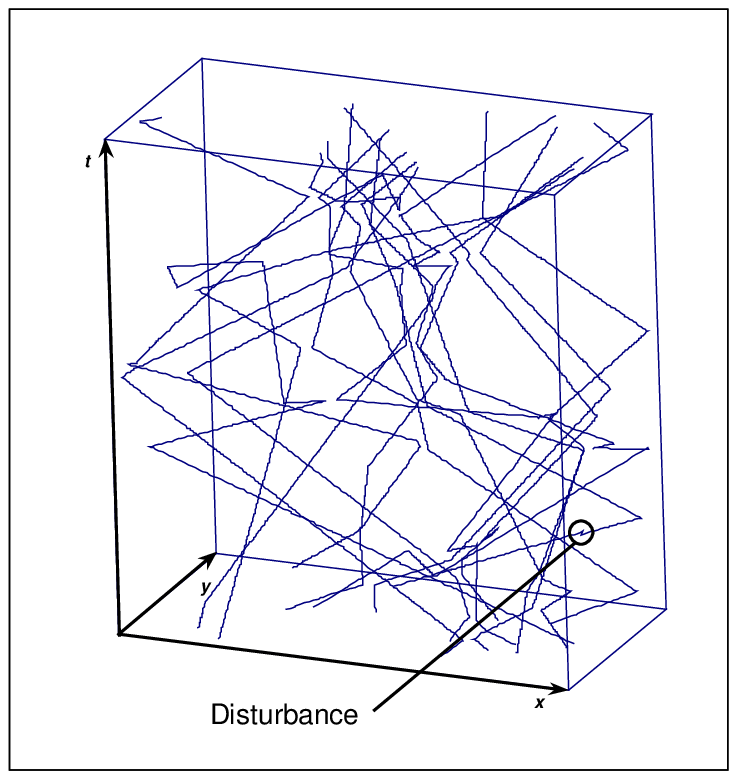}
\renewcommand{\thefigure}{4b}
\caption{The same computer simulation as in 3b, with a similar
disturbance.
Here, the return to the ordered initial state fails.}
\end{center}
\end{figure}

Fig.~3 gives a simple illustration for this principle. A set of Newtonian
balls (hence, harboring an $\bf I$ topology) is simulated in the forward and backward time
directions at absolute precision. In Fig.~4, an event of $\bf X$ indeterminism is
inserted by a single random interference with the position of a single
ball. It is common knowledge that, for a normal system whose entropy
increases, a slight interference will make no difference for the overall
entropy increase. The final macrostate affected by the interference will
be indistinguishable from the final state that would ensue otherwise. In
contrast, an entropy {\it decreasing} system is extremely sensitive to
interference: It requires a very special microstate at any instant, and
the slightest interference will ruin its entropy decrease. Hence, the
system will fail to converge to the desired ordered state.

What makes this seemingly trivial observation crucial in the context of
time's arrow is this: Regardless of the system's initial conditions, its
entropy {\em always} increases following the random event. If these
initial conditions are of the normal, entropy-increasing type, the
random event would merely strengthen entropy increase. If, on the other
hand, they are of the unique, entropy-{\em decreasing} type, the random
event would most probably flip the system's evolution back towards
entropy increase. Hence, genuine indeterminism at the micro scale
enforces a macroscopic time arrow that has nothing to do with initial
condition. Rather, it seems to be intrinsic (see also Elitzur and Dolev
\cite{ED1999a,ED1999b,ED2000}).

A similar conclusion has been reached by Albert
\cite{Alb1994a,Alb1994b}, who studied the indeterminism entailed by the GRW theory. We shall review the GRW interpretation in the following section, and Albert's arguments afterwards.

\section{The GRW theory}
\label{sec:grw}

The GRW theory \cite{GRW1986} is an interpretation of QM
in which the von Neumann collapse is built into the dynamics of the
wavefunction itself.
There is no talk about measurement, observation, decoherence, or
anything like that, and there is no need for thumb rules that somehow
distinguish between the classical and the quantum levels, or system and
its
environment. According to GRW, any system whatsoever has a quantum
state, and
that state evolves under a single dynamical equation of motion.
Quantum mechanics can be applied to any system whatsoever, including the
universe as a whole. However, the theory explains, as a straightforward
result of the dynamics, why systems consisting of a large number of
particles (\ie macroscopic systems) do not usually exhibit quantum
interference (Schr\"odinger cats and measuring devices included).
The theory in its most simple (and nonrelativistic) form goes as
follows.\footnote{Here we follow Bell's discrete model \cite{Bel1987}.}

Consider the quantum wavefunction of a composite system
consisting of $N$ particles:
\begin{equation}
\ket{\Psi(x_1...x_N,t)}.
\label{eq:psi}
\end{equation}
The time evolution of the wavefunction {\em usually but not
always} satisfies the linear and deterministic Schr\"odinger
equation. From time to time the
wavefunction {\em collapses} onto some {\em localized Gaussian in
position} of the normalized form:
\begin{equation}
    G(x_k)=Ke^{-(x-x_k)^2/2\Delta},
\label{eq:gau}
\end{equation}
where $x_k$ is randomly chosen from the arguments in (\ref{eq:psi})
and the width $\Delta$ of the Gaussian is stipulated to
be approximately $\Delta=10^{-5}cm$. This parameter is taken as
a new constant of nature. The GRW jumps are also
stipulated to occur only in {\em position}, and this means that
on this theory, wavefunctions that are approximate eigenstates of
position are taken to be physically {\em preferred}.

Two additional questions with respect to the GRW collapses need to
be settled: first, When do they occur? and second, Where?
According to GRW, the answers to both questions are a matter of
{\em probability}. Concerning the first, the probability $\tau$ for
a collapse for a single particle at any given time is stipulated to be
approximately $\tau=10^{-15}$ (again a new constant
of nature). Concerning the second question, the probability
that the reduced wavefunction will be centered around any
spatial point $x$ at time $t$ is given by the
standard Born rule:
\begin{equation}
  \abs{\braket{\Psi(x_1...x_N,t}{G(x_k)}}^2,
\label{eq:born}
\end{equation}
where $G(x,k)$ is the GRW Gaussian defined in (\ref{eq:gau}).

This is essentially the whole theory. However, three brief comments are in order:

(1) The choice of the new constants of nature, \ie the frequency of
the collapses $\tau$ (which is proportional to the number of
particles) and the width of the Gaussian $\Delta$ ensure that
the collapses will not result in violations of either our experience
on the macroscopic scale, or of the well confirmed predictions of
standard
quantum mechanics. Moreover, the stochastic dynamics is designed in such
a way that the solution to the measurement problem in the quantum theory of
measurement, obtains a straightforward result of the quantum mechanics
of
composite systems. In other words, any
interaction which involves the {\em spatial displacement} of a large
number of particles (say, in the pointer, or in records imprinted on a
paper) will, with overwhelming probability, result in a collapse of the
wavefunction onto one of the eigenstates of the pointer's position with
the usual Born probabilities.

(2) It follows from the GRW
prescription that the collapses of the wavefunction might induce
violations
of both momentum and energy conservation. The GRW collapses will
sometimes give rise to increase in momentum, in which case
electrons might jump out of their orbits in the atoms. Similarly, they
might
increase the energy of a given cloud of gas just enough to
heat it up, whereas we know experimentally that such things
do not occur. However, $\Delta$ is chosen here to be wide enough so that
the violations of the conservation laws will be small enough so as
to be unobserved. So (in principle) it seems as if in the GRW
theory there is a trade off between getting the {\em second} law of
thermodynamics right, by the price of making the {\em first} law approximate. Note that if the
GRW collapses were to occur onto {\em delta} functions in position, the
violations of the conservation laws would be observable.

(3) $\Delta$ is chosen to be narrow enough so that the
GRW collapses map the wavefunctions onto states that are
close (in Hilbert space norm) to eigenstates of position, so that the positions become only
approximately definite. This means that the GRW hits do not set the
systems onto definite position states,
but rather onto superpositions of positions with nonzero
tails.\footnote{This yields the so-called tails problem: see Albert \cite{AL1990}.
Also, when and where the hits occur in measurement
situations might drastically vary. For some measurements it turns out
that the GRW collapses are unlikely to occur prior to the interaction with
the observer: see Albert and Vaidman \cite{AV1989}.} But the
amplitudes of these tails are small enough, so that the superposition has
no observable effects, and the particles behave
as if they actually have definite positions. Moreover,
it turns out that these tails play
an important role in the attempts to write down a relativistic
version of the GRW theory.\footnote{\label{note:GRW} In the case of the continuous, relativistic, model of
the GRW theory the time reversed evolution in 
standard quantum mechanics (with no collapse) and in the GRW theory
coincide. That is, on this model the almost zero tail of
the "down" wavepacket will grow as the wavefunction is evolved backwards 
in time, so that the reversed wavepackets will exhibit the standard
Schr\"odinger re-interference. This is by contrast to the almost
negligible effect of the tail of the wavefunction in the course of the
forward time evolution. In this sense also the continuous model of the
GRW theory is not time reversible, though for a different reason. Note
that the time irreversibility of the GRW tails turns out to be crucial in
the case of relativistic versions of the theory. See Pearle
\cite{Pearle1989}. Given this behavior of the continuous model, it 
is interesting to consider whether or not it remains genuinely
stochastic.}

\section{Albert's Derivation of Entropy Increase from Quantum Collapse}
\label{sec:alb}

According to Albert \cite{Alb1994a,Alb1994b}, the GRW theory  provides a
good candidate for reducing thermodynamics to mechanics, because
its basic {\em dynamical} laws are fundamentally and irreducibly
{\em indeterministic}. Moreover, the genuine stochastic collapse in the
GRW theory is supposed to occur at the {\em microscopic} level with
a frequency that suffices to actually
{\em derive} the second law from the GRW dynamics.

Albert argues that
a fundamental thermodynamic arrow intrinsic to the dynamics can be
derived only in the GRW theory and not in other interpretations of
quantum mechanics, such as Bohm's theory, modal interpretations and the
many
worlds interpretations, since the latter are all deterministic and time
reversible theories ($\bf I$ topology). Following Albert, we argue that a genuine indeterministic dynamics is a
{\em necessary} (though not sufficient) condition for deriving a
fundamental arrow of time, including the thermodynamic arrow. That is,
such a derivation can be carried only in theories which
employ genuine stochastic equations of motion.
By an {\em intrinsic arrow of time}
we mean that the arrow of time should be solely a consequence of the
{\em dynamical} equations of motion of the theory, and in
particular it should be independent of initial (or final) conditions.

Here is one way to see how Albert's reduction of the thermodynamic laws to
the GRW theory is carried out. Consider, for instance, the special case of  
the approach to equilibrium of thermodynamic systems. 
Take $N$ molecules of some gas that is spreading out in a container.
In standard (Schr\"odinger) quantum mechanics the composite wavefunction
of the gas will almost always correspond to a state in which (due to
quantum mechanical interactions) the molecules will be entangled with
each other (and with the container walls), and moreover
they will generally not be located in some definite positions
within the container.

However, because of the high value of N, the $N$-particles wavefunction in the GRW theory has an
overwhelming probability for collapses at almost every
instant of time. Moreover, according to the GRW prescription
the wavefunction is reduced after a collapse to a Gaussian that is
localized around some {\em spatial} distribution of the gas
molecules. This means in particular that the solutions of the GRW
equations of motion for the times at which collapses occur are
approximately product states in {\em position}, where the molecules
have in fact definite positions.

Finally note that
this behavior critically depends on two factors. First,
the number of particles in the system needs to be large enough to
ensure the high probability for a GRW collapse.
Second, the time evolution needs to be such that the wavefunction
evolves
into a superposition of terms corresponding to spatial locations of
the molecules which differ more than the width $\Delta$ of the GRW
Gaussian. If these conditions hold, the GRW jumps will invariably
change the wavefunction of the gas in a way that is enough to
put each molecule of the gas onto an apparently well
defined trajectory. In particular, {\em this behavior of the
wavefunction
is completely independent of initial conditions.}

Recall now how mainstream physics tackles Loschmidt's paradox.
According to Loschmidt, by the same statistical
considerations that have lead Boltzmann to predict entropy growth in the
future direction, entropy should grow in the {\em past} direction too, as
statistics itself is indifferent to the time directions. Against this
paradox physicists invoke the
{\em low entropy past postulate} (see Price \cite{Pri1996}). This
hypothesis simply imposes {\em by fiat} the low-entropy state in the
universe's past, in accordance with everyday observation. However, in the
case of completely deterministic and time reversible theories this
hypothesis means that the thermodynamic laws are recovered only for very
special initial conditions which are (moreover) highly {\em
improbable}.

In the case of the GRW theory, on the other hand, Loschmidt paradox simply does not
arise. That is, the dynamics by itself does not entail that low
entropy states in the past are just as improbable as low entropy
future states. In this case, in order to settle the question whether or
not entropy increasing trajectories in the future direction 
are highly probable, one has to actually {\em solve} the equations of
motion of the GRW theory for the time in question.
And that, as repeatedly stressed by Albert, is a
straightforward empirical question (though perhaps not a tractable
one in practice) about the {\em predictions} of the
theory.

 That is, given the GRW jumps, the question of the
thermodynamic
arrow of time becomes a completely empirical question.
What is clear is that in the GRW dynamics there is a built in
mechanism in the form of the quantum jumps that as a matter of
principle can send the system onto an entropy increasing trajectory
independently of initial conditions, in a way similar to our simulation on Section 3. And this is ensured by the
fact that the GRW dynamics is truly {\em stochastic}
(or indeterministic), and
a-fortiori time irreversible.\footnote{There are some extreme cases in
which, as of principle, the GRW theory might not deliver the second
law. But due to decoherence effects these are probably unobservable,
see Albert \cite{Alb1994a}.}

\section{The Arrow of Time}
\label{sec:arrow}

We conclude our paper by generalizing our argument (Elitzur and Dolev
\cite{ED1999a,ED1999b,ED2000}): A real arrow of
time, that is an arrow of time intrinsic to the dynamical
equations of motion, exists only in theories with indeterministic
dynamics. In this sense, we argue that indeterministic $\bf X$ topology dynamics,
as in the GRW theory, is a {\em necessary} (though not always
sufficient) condition for deriving the thermodynamic arrow of time.

As is well known the postulated dynamics in most theories of current
physics is time reversible.\footnote{Quantum weak interaction
with $CP$ violations notwithstanding.} This means that in principle such theories
cannot
account for the thermodynamic arrow of time (and for our experience of
the arrow of time\footnote{We assume here that our experience of
the arrow of time supervenes on physical processes.})
as intrinsic to the dynamical equations of motion. As we briefly discussed in section
\ref{sec:alb}, such theories can recover the second law of
thermodynamics
only by appealing to the low-entropy past hypothesis.
This applies to both classical statistical mechanics and to no-collapse quantum mechanics (including decoherence
theories).\footnote{For how to recover the thermodynamic laws in no
collapse quantum mechanics, see Hemmo and Shenker \cite{HS2001}. It
turns out, however, that environmental decoherence is a sufficient
condition for deriving the thermodynamic arrow as an effective law.}
In such theories no arrow of time is built into the dynamics (since they
are time reversible), and therefore no {\em intrinsic} arrow of time can
be obtained from the dynamics, no matter what empirical
evidence we have for the thermodynamic arrow. This is
not the case in stochastic theories such as the GRW theory.

Thus it is crucial to bear in mind that even if the second law of
thermodynamics (and thus the thermodynamic arrow of time) can
be derived in completely deterministic and time reversible theories,
this would mean that the thermodynamic behavior
is recovered only as an {\em effective} behavior that crucially 
depends on some specific initial conditions. In this sense such behavior
will not be intrinsic to the dynamics of the theory. 
The microcanonical distribution postulated by the traditional
approaches to this problem in classical statistical mechanics, for
instance, ensures that the second law can be derived as an effective
law. But the microcanonical distribution itself does
not depend on the dynamics,\footnote{There is a sense in which
the ergodic approach in classical statistical mechanics tries to derive
the microcanonical distribution from the ergodicity of the dynamics. But
it is not clear whether this
approach can be successful. See Sklar \cite{Sklar1993} for an overview 
of the problem and of the ergodic approach, and for references.} and so
thermodynamically pathological trajectories are in principle possible
depending solely on the initial conditions of the system. Moreover,
such trajectories turn out to be improbable, that is, they are
assigned low probability (whatever probability means in the
classical approach) only on the low-entropy past hypothesis.

Note that a theory might be genuinely stochastic and yet not deliver the
thermodynamic arrow. For to derive the latter the stochastic dynamics 
at the macroscopic level must also be time irreversible, and (moreover)  
in the case of quantum mechanics the stochastic jumps must result
in approximate eigenstates of {\em position}. That is, a stochastic theory
in which the jumps were to occur, say onto some eigenstates of momentum, 
will definitely not yield the thermodynamic arrow. Since the above
additional conditions on the dynamics are independent of whether or not
the dynamics is  stochastic, it follows that indeterminism by itself, even
if it contains a built in arrow of time, is not sufficient to derive the thermodynamic
arrow. This is, in fact, why other collapse
theories such as von Neumann's standard formulation will not
be workable for deriving the arrow of time in general.\footnote{In
von Neumann's theory the stochastic dynamics comes to play only in 
measurement situations, and so the derivation will only hold for these
situations. This means that on a such a theory there will be no
{\em universal} thermodynamic arrow, since no measurements are carried 
out on the universe as a whole.}

In the case of the GRW theory, the
quantum jumps are not time reversible, and that very fact allows to define, what might be called a {\em quantum mechanical arrow of
time}. As we see
from the above example, however, this arrow is completely independent
of the thermodynamic arrow. But in the case of the GRW
theory we saw that the dynamics alone can yields {\em also} the
thermodynamic arrow of time. In this sense, in the GRW theory the
thermodynamic arrow correlates with the quantum mechanical arrow of
time. Moreover, if, as a matter of empirical fact, the GRW equations of
motion will turn out to yield, with high probability, entropy increasing
trajectories, then one could say that on this theory the thermodynamic
arrow is {\em derived} from the quantum mechanical arrow.
That is, what we have here is a proof that in the GRW theory the
thermodynamic arrow can be built from the specific form of the GRW
jumps.

\bibliographystyle{unsrt}

\end{document}